\documentclass[conference]{IEEEtran}
\IEEEoverridecommandlockouts
\usepackage{cite}
\usepackage{amsmath,amssymb,amsfonts}
\usepackage{algorithmic}
\usepackage{graphicx}
\usepackage{textcomp}
\usepackage{xcolor}
\usepackage[a4paper, total={184mm,239mm}]{geometry}
\usepackage{multirow} 
\usepackage{booktabs} 
\usepackage[ruled,linesnumbered]{algorithm2e}
\usepackage{listings}
\usepackage{enumitem}
\usepackage[T1]{fontenc}
\newcommand{\ra}[1]{\renewcommand{\arraystretch}{#1}}
\newcommand{\N}{SuperFlow}

\newcommand{\affmark}[1]{\textsuperscript{#1}}
\def\BibTeX{{\rm B\kern-.05em{\sc i\kern-.025em b}\kern-.08em
    T\kern-.1667em\lower.7ex\hbox{E}\kern-.125emX}}
\begin{document}

\title{SuperFlow: A Fully-Customized RTL-to-GDS Design Automation Flow for Adiabatic Quantum-Flux-Parametron Superconducting Circuits
\thanks{*Equal contributions.}
}

\author{
    \IEEEauthorblockN{
        Yanyue Xie\affmark{1}*, Peiyan Dong\affmark{1}*, Geng Yuan\affmark{2}, Zhengang Li\affmark{1}, Masoud Zabihi\affmark{3}, Chao Wu\affmark{1}, Sung-En Chang\affmark{1}, \\
        Xufeng Zhang\affmark{1}, Xue Lin\affmark{1}, Caiwen Ding\affmark{4}, Nobuyuki Yoshikawa\affmark{5}, Olivia Chen\affmark{6} and Yanzhi Wang\affmark{1} \\
    }
    \IEEEauthorblockA{
        \textit{\affmark{1}Northeastern University, \affmark{2}University of Georgia, \affmark{3}IBM Research,}\\
        \textit{\affmark{4}University of Connecticut, \affmark{5}Yokohama National University, \affmark{6}Tokyo City University}\\
        \affmark{1}\{xie.yany, dong.pe, yanz.wang\}@northeastern.edu, \affmark{6}olivia.chen@ieee.org
    }
}

\maketitle

\begin{abstract}

Superconducting circuits, like Adiabatic Quantum-Flux-Parametron (AQFP), offer exceptional energy efficiency but face challenges in physical design due to sophisticated spacing and timing constraints. Current design tools often neglect the importance of constraint adherence throughout the entire design flow.
In this paper, we propose \N{}, a fully-customized RTL-to-GDS design flow tailored for AQFP devices.
\N{} leverages a synthesis tool based on CMOS technology to transform any input RTL netlist to an AQFP-based netlist.
Subsequently, we devise a novel place-and-route procedure that simultaneously considers wirelength, timing, and routability for AQFP circuits.
The process culminates in the generation of the AQFP circuit layout, followed by a Design Rule Check (DRC) to identify and rectify any layout violations.
Our experimental results demonstrate that \N{} achieves 12.8\% wirelength improvement on average and 12.1\% better timing quality compared with previous state-of-the-art placers for AQFP circuits.


\end{abstract}

\section{Introduction}
\label{sec:introduction}


Superconducting logic circuits exhibit extremely high energy efficiency over their Complementary Metal–Oxide–Semiconductor (CMOS) counterparts \cite{takeuchi2013adiabatic}.
Adiabatic Quantum-Flux-Parametron (AQFP) logic \cite{harada1987basic} is a representative energy-efficient superconducting logic that is designed to achieve a reduction in both static and dynamic power consumption by adopting adiabatic switching \cite{takeuchi2014energy}.
AQFP can potentially achieve $10^4-10^5$ energy efficiency gain compared with state-of-the-art CMOS technology with a clock frequency of several GHz \cite{cai2019stochastic}.
AQFP differs from CMOS circuits in terms of active components, passive components, logic gates, data propagation, clocking scheme, fan-out requirements, and power consumption, as detailed in Table \ref{tab:comparison}. Therefore, the design automation tools developed for CMOS cannot be directly applied to the design of AQFP circuits.

\begin{table}[!htbp]
\centering
\ra{1.1} 
\caption{Comparison of AQFP with CMOS.}
\label{tab:comparison}
\resizebox{1.0\columnwidth}{!}{
\begin{tabular}{l|ll}
\toprule
\textbf{Circuits}          & \textbf{AQFP}             & \textbf{CMOS} \\ \midrule
\textbf{Active component}  & Josephson junction (JJ)   & Transistor    \\ \hline
\textbf{Passive component} & Inductor                  & Capacitor     \\ \hline
\textbf{Logic gate}        & Majority-based gates      & And, or, inverter gates  \\ \hline
\textbf{Data propagation}  & Current pulse             & Voltage level \\ \hline
\textbf{Clocking}          & Four-phase clocking       & Synchronous   \\ \hline
\textbf{Fan-out}           & $=1$(Splitter for fan-outs) & $\ge 1$    \\ \hline
\textbf{Power}             & Alternating Current (AC)  & Direct Current (DC) \\ 
\bottomrule
\end{tabular}
} 
\end{table}

There have been several existing works on customized design automation tools for AQFP circuits.
\cite{huang2021optimal,cai2019majority,lee2022beyond} solve the buffer and splitter insertion problem during the logic synthesis stage for path balancing and fan-out branching imposed by AQFP technology.
\cite{li2021towards} and \cite{chang2020asap} consider spacing constraints and max-wirelength constraints of AQFP during the placement stage.
\cite{dong2022taas} further improves the timing of AQFP designs.
However, all of these works focus solely either on the logic synthesis stage or the placement stage, without addressing the entire design flow, i.e., a complete RTL-to-GDS (Register-Transfer Level to Graphic Design System) flow.
While some studies offer complete design flows, they only focus on specific steps like logic synthesis or AQFP cell library design~\cite{meuli2022majority}~\cite{tanaka2023full}, both leaving placement and routing stage to commercial tools and lacking flexibility.

Transitioning from CMOS to the more energy-efficient AQFP circuits necessitates a comprehensive and dedicated AQFP design automation tool flow to optimize Power, Performance, and Area (PPA).
The absence of such a comprehensive tool flow restricts design optimization, potentially leading to issues like congestion and post-routing timing violations.
Standard electronic design automation (EDA) tools for CMOS logic are not applicable to AQFP logic, due to inherent differences such as clocking constraints and fan-out requirements. 
Relying on commercial EDA tools for certain or all stages would result in limited flexibility, particularly considering that AQFP is an emerging technology and the AQFP cell library is under active development \cite{tanaka2023full}. 
In light of these challenges, we develop a fully-customized RTL-to-GDS design automation flow tailored for AQFP circuits, enabling designers to easily adjust the design objectives for AQFP and incorporate timely updates to the AQFP cell library.

Our major contributions can be summarized as follows:
\begin{itemize}
    \item We present a fully-customized RTL-to-GDS design flow, \N{}, for AQFP circuits. To the best of our knowledge, this is the first non-commercial RTL-to-GDS design automation tool that targets AQFP devices. 
    \item We optimize the placement quality by optimizing wirelength and timing simultaneously while respecting the clocking and the mixed-cell-size constraints at the detailed placement stage. 
    \item We introduce a layer-wise routing strategy, capable of routing with space expansion, thereby addressing potential routability issues.
    \item Experimental results show that \N{} achieves 12.8\% wirelength improvement on average with 12.1\% better timing quality over previous state-of-the-art placers for AQFP circuits. 
\end{itemize}

\section{Background}
\label{sec:background}

\begin{figure}[htbp]
    \centering
    \includegraphics[width=1.0\linewidth]{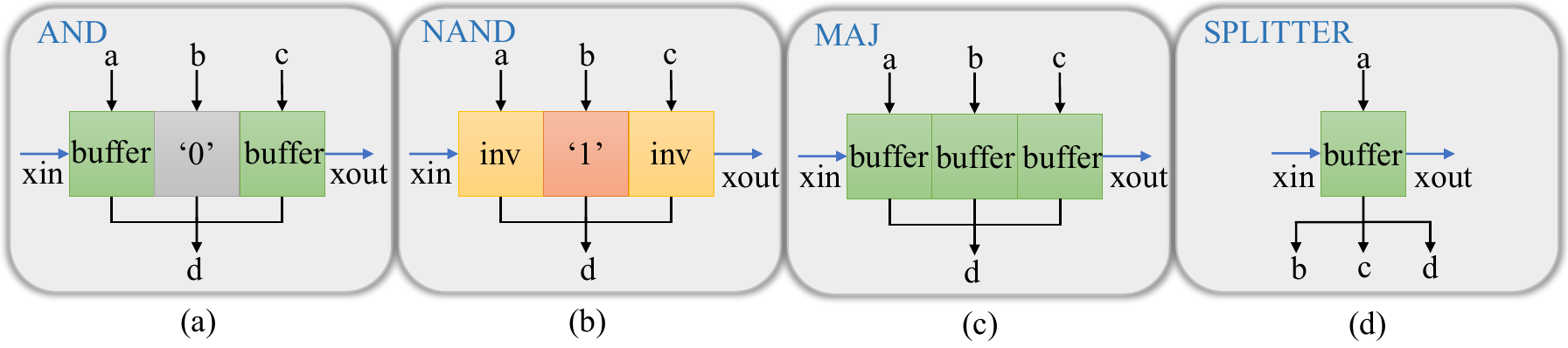}
    \vspace{-0.7cm}
    \caption{\textbf{The symbols of AQFP logic gates.} (a) \texttt{and} gate; (b) \texttt{nand} gate; (c) \texttt{majority} gate; (d) \texttt{splitter} gate.} 
    \label{fig:aqfp_cells}
\end{figure}

\vspace{-0.4cm}
\subsection{AQFP Superconducting Logic}
\label{subsec:aqfp_cells}
The fundamental building block of AQFP logic gates is the AQFP \texttt{buffer}, which consists of a double-Josephson junction (JJ) superconducting quantum interference device (SQUID) \cite{takeuchi2013adiabatic}. Together with the inverter gate and constant gate, these basic gates constitute the AQFP cell library, including 
\texttt{and}, \texttt{or}, \texttt{not}, \texttt{majority}, \texttt{buffer}, and \texttt{splitter}, as shown in Fig.~\ref{fig:aqfp_cells}. The AQFP standard cell library is built via the minimalist design approach
, utilizing a bottom-up method to construct more complex gates~\cite{cai2019majority}. The design of the AQFP cell library follows the AIST standard process 2 (STP2) and the MIT Lincoln Laboratory (MIT-LL) SQF5ee fabrication process, both of which are niobium-based integrated-circuit technologies suitable for AQFP.

Different from traditional CMOS logic that utilizes and-or-inverter-based (AOI) representation, AQFP logic favors majority-based (MAJ) gates, due to the efficient utilization of JJ resources. Three-input majority gates in AQFP consume the same amount of JJ resources as two-input AOI gates~\cite{cai2019majority}. Please note that, unlike CMOS gates which can connect to fan-out directly, all AQFP gates require the use of \texttt{splitter} cells for multiple fan-outs.


\subsection{AQFP Clocking Architecture}
\begin{figure}[htbp]
    \centering
    \includegraphics[width=0.7\linewidth]{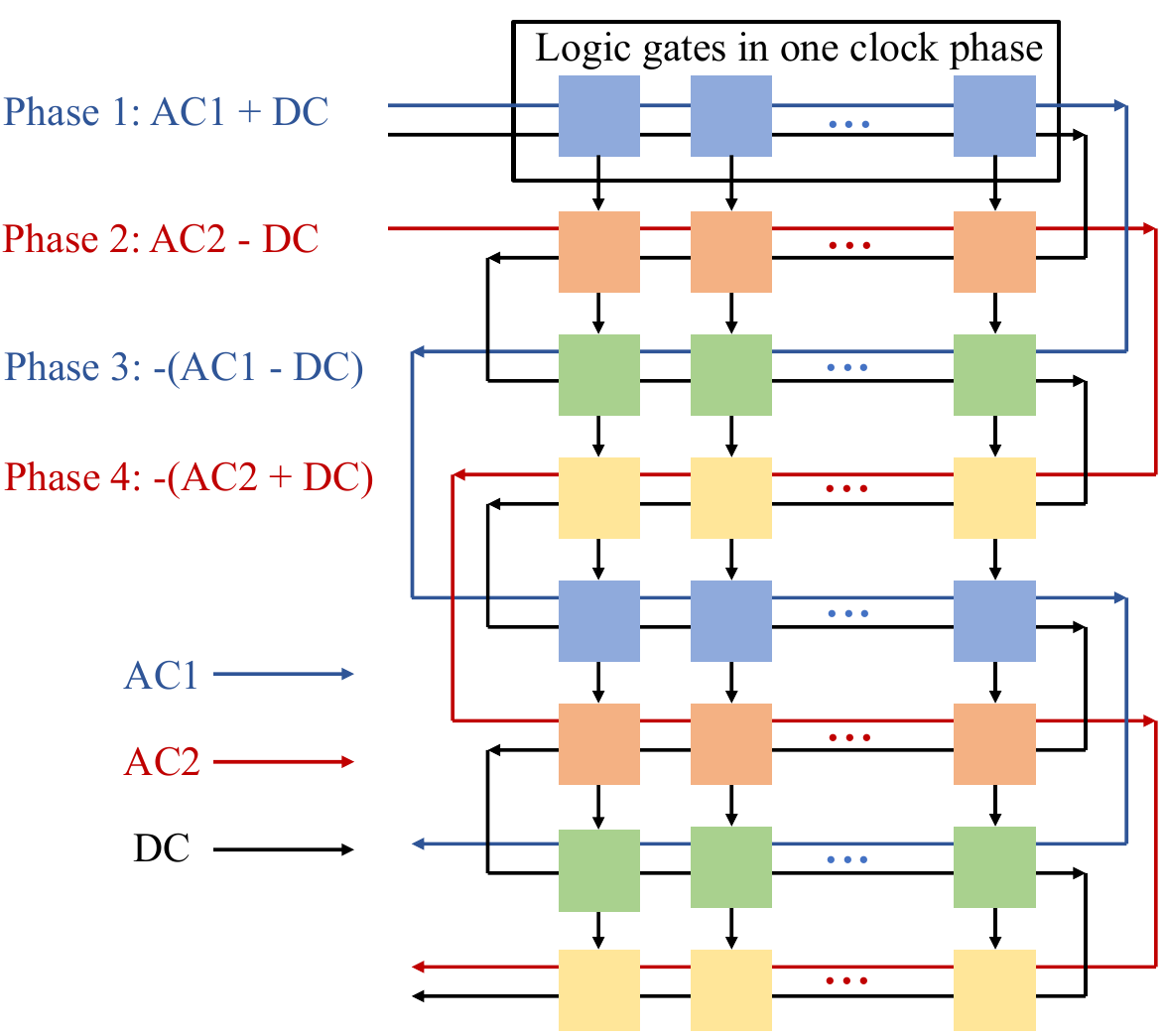}
    \vspace{-0.3cm}
    \caption{\textbf{AQFP clocking architecture.} AQFP clocking scheme utilizes one DC and two AC signals to create a four-phase clocking scheme. The two AC signals have a phase difference of 90 degrees. The intricate zigzag clocking signals impose strict timing constraints on AQFP logic cells and formulate a deep-pipelined architecture.}
    \label{fig:aqfp_clocking}
\end{figure}

The clocking architecture and timing requirements of AQFP differ from CMOS.
In CMOS circuits, multi-level gates are expected to meet timing constraints as a group, i.e., multi-level gates for pipelining, while in AQFP circuits, each gate must satisfy the timing requirements, i.e., the gate-level pipelining.
Specifically, four-phase AC bias currents serve as both power supply and clock signal~\cite{takeuchi2013adiabatic}, which triggers the data from one clock phase to the next, as illustrated in Fig.~\ref{fig:aqfp_clocking}. 
AQFP utilizes this feature to mitigate the power consumption overhead of DC bias shown in other superconducting logic technologies, such as RSFQ~\cite{likharev1991rsfq}. 
While AC biasing contributes to exceptional energy efficiency, AQFP circuits also adopt a deep-pipelined architecture since each AQFP logic gate is connected to an AC clock signal and occupies one clock phase.
This deep-pipelined architecture mandates all inputs for a logic gate to have the same delay (clock phases) from the primary inputs \cite{cai2019majority}, necessitating rigorous path balancing.
Furthermore, when it comes to a complete design flow, AQFP's gate-level pipelining provides more flexibility, allowing simultaneous optimization of spacing and timing constraints.

\vspace{-0.3cm}
\subsection{Challenges of AQFP Physical Design}

AQFP physical design is complicated and challenging due to various design constraints:
(\romannumeral1) The AQFP placement problem involves fulfilling multiple spacing limitations, including cell spacing and zigzag spacing constraints~\cite{li2021towards}. Cell spacing requires that neighboring cells in a row either touch or maintain a minimum distance. Regarding zigzag spacing, when vias are utilized to change the wire directions, the wire zigzags must adhere to a predetermined minimum spacing (e.g., 10$\mu$m for the MIT-LL process).
(\romannumeral2) AQFP circuits should comply with maximum wirelength requirements, which specify that a single wire connection should not exceed $W_{max}$ \cite{chang2020asap}. If a single connection exceeds the maximum wirelength restriction, it is necessary to insert an entire row of buffers between the two rows.
(\romannumeral3) Due to the different sizes of AQFP buffers relative to other majority-based cells, and the possibility of having various types of cells within a single clock phase, AQFP placement presents a mixed-cell-size problem. Within a single densely populated clock phase with multiple mixed-cell-size cells, perturbations or cell swapping can potentially cause significant intra-row violations.
(\romannumeral4) Strict timing constraints should be maintained since AQFP circuits run at a clock frequency of several GHz. The zigzag clocking scheme and the deep-pipelined architecture impose great challenges to AQFP placement algorithms as the cell positions in a fixed row have a significant impact on the timing closure. As a result, traditional placers \cite{li2021towards, chang2020asap} that address wirelength alone may lead to placement results with severe timing violations.
(\romannumeral5) AQFP routing is limited to two metal layers between two adjacent clock phases, conforming to the zigzag clocking architecture and process requirements.

Hence, relying solely on separate synthesizers, placers, or routers for AQFP circuits is insufficient. A comprehensive design flow that encompasses and adheres to the various design constraints of AQFP circuits is imperative. For instance, following logic synthesis and the insertion of buffers/splitters for path balancing, each logic cell is assigned a specific clock phase or row index. The placement stage should preserve the assigned row index for each logic cell while optimizing their horizontal positions within the rows. Subsequently, during the routing stage, clock wires must be routed based on the placement results, ensuring no violations occur. Furthermore, both placement and routing must consider clocking constraints to achieve optimal timing and meet the desired clock frequency.

\section{Methods}
\label{sec:methods}

\subsection{Overall AQFP Design Flow}
\begin{figure}[htbp]
    \centering
    \includegraphics[width=0.8\linewidth]{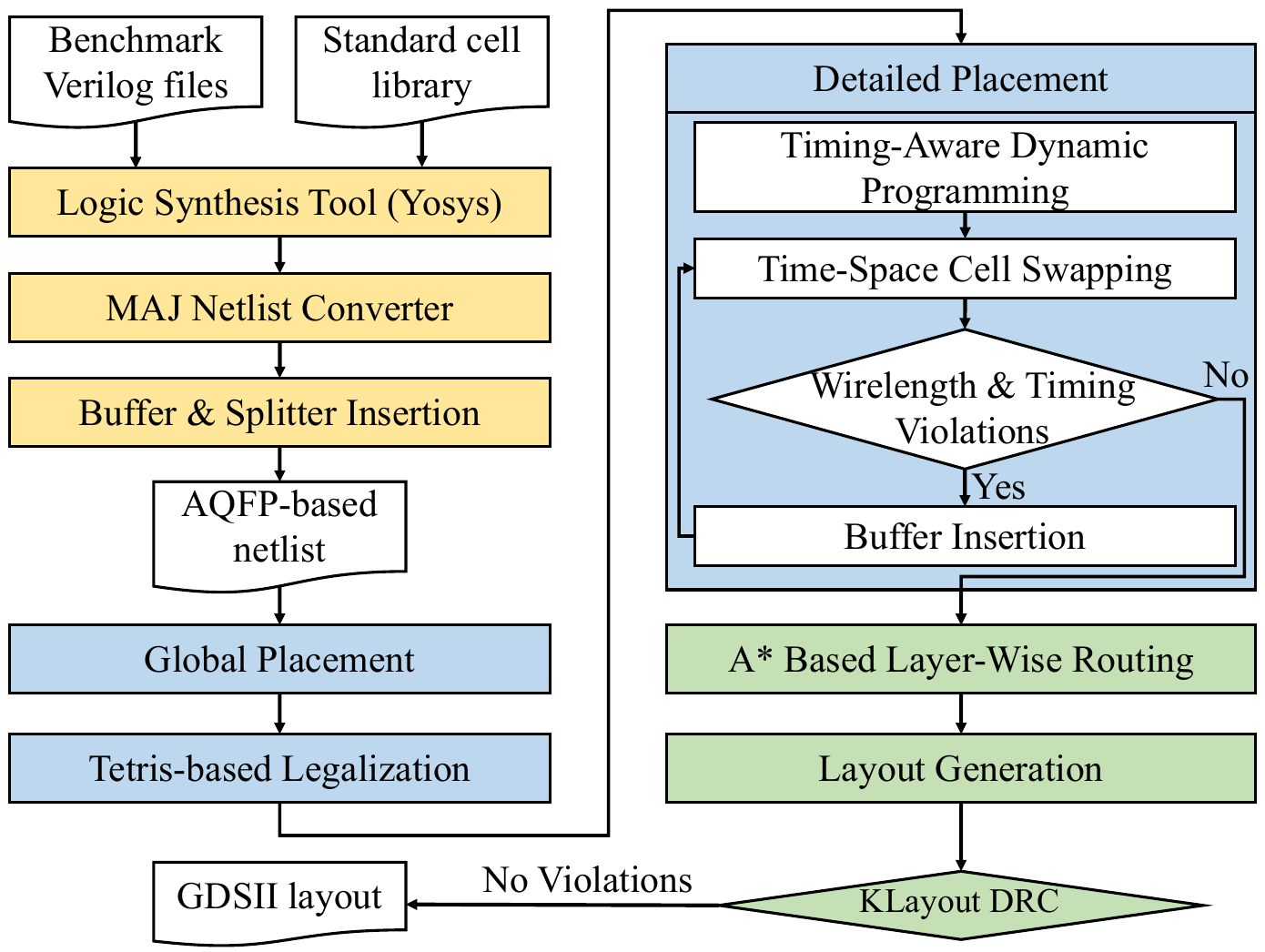}
    \vspace{-0.3cm}
    \caption{\textbf{The overall design flow of \N{}.}}
    \label{fig:aqfp_flow}
\end{figure}

Fig.~\ref{fig:aqfp_flow} demonstrates the overall design flow of \N{} for AQFP circuits.
The input files are the standard cell library of AQFP and the Register-Transfer Level (RTL) files describing circuit architecture.
After logic synthesis and an AQFP interpreter, \N{} maps the netlist to an AQFP-based netlist, which serves as the input to the following place-and-route step.
Then \N{} generates the circuit layout based on the physical information provided after the place-and-route step.
Should any violations emerge after the Design Rule Check (DRC), \N{} automatically rectifies these violations and proceeds to finalize the layout files for AQFP circuits in Graphic Design System II (GDSII) format

\subsection{Majority-Based Logic Synthesis for AQFP}

\subsubsection{\textbf{Majority Netlist Conversion}}
Fig.~\ref{fig:aqfp_flow} illustrates the logic synthesis stage. AQFP prefers a majority-based netlist and we first leverage a CMOS-based synthesis tool, Yosys, to generate AOI netlists for input RTL files.
The AOI netlist is then converted to an MAJ netlist by mapping all feasible three-input nets to their corresponding two-level majority-based implementation while maintaining the minimum overall resource consumption. 

To elaborate, we view the netlist as a directed graph. In an AOI netlist, each node has two parent gates. While in a majority-based netlist, each node has two or three parents depending on its type of logic gate, e.g. an \texttt{and} gate or a \texttt{majority} gate. The majority netlist conversion involves three steps: 
(\romannumeral1) Identifying convertible three-input nets in AOI netlists. (\romannumeral2) Applying a table-based method to map these nets to majority-based logic. (\romannumeral3) Selecting the most resource-efficient mapping for each net.


We initiate by identifying feasible three-input nets in the AOI netlist using a depth-first search, starting from the netlist output. The search begins with a two-input net formed by the node and its two parent inputs, progressively adding parent nodes to expand the net. This iterative search continues until three independent parent nodes are found, ensuring no parent node is a descendant of another in this search. If more than three parent nodes are identified, the search is aborted, deeming the node unsuitable for a three-input net.

Feasible three-input nets identified earlier can be mapped to majority-based gates. Each net can be represented by up to two majority gates. To check for majority mapping suitability, we use a Karnaugh map checking method. This method compares the Karnaugh maps of all combinations of three parent nodes with the target net. If a match is found with primary gates, the net is mapped directly; otherwise, it is mapped to two-level majority-based logic with three majority gates at the first level and one at the second level, connecting to the outputs of the first three gates.

In optimizing the majority-based netlist, our goal is to minimize both Josephson junction count and clock delays. This requires a thorough search across the entire directed graph. We prioritize majority mappings that use more logic gates and clock phases, iterating through all mapping schemes for each three-input net to find the one that uses the least resources overall.

\subsubsection{\textbf{Buffer and Splitter Insertion}}
After converting the AOI netlist to a majority-based netlist, we insert \texttt{splitter} cells to the converted netlist to comply with the AQFP fan-out requirements.
Gates with over two fan-outs require a \texttt{splitter} cell, selected based on the number of fan-outs.
After addressing all fan-outs, we reset and recalculate net delays.
Given the updated delay for each net, buffers are inserted in each data path to every gate accordingly to produce equal delay (clock phase), which is required by the AQFP technology \cite{cai2019majority}.
Since the logic structure of the converted netlist is fixed, buffer insertion could be resolved in any order and will not change the total clock phases or the critical path.
With all buffers and \texttt{splitter} cells in place, we finalize the majority-based AQFP netlist for subsequent placement and routing stages.

\subsection{AQFP Placement}
The placement in Fig.~\ref{fig:aqfp_flow} can be divided into global placement, legalization, and detailed placement. In this section, we clarify the implementation details of these steps.

\subsubsection{\textbf{Problem Formulation}}

The AQFP placement problem can be formulated as a hypergraph $G = (V,E)$, where $V$ and $E$ denote the cells and nets, respectively. Let $x_i$ and $y_i$ be the $x$ and $y$ coordinates of the center of logic cell $v_i$. Given the AQFP clocking architecture and a design netlist, our objective is to determine the positions of the cells that optimize the routed wirelength subject to the following constraints:
\begin{itemize}
    \item Each logic cell should maintain consistent delay (clock phase), respecting the sequence from the netlist.
    \item Cells must not overlap with others, and two horizontally neighboring cells in a row can either be abutting or keeping a minimum spacing $s_{min}$ (spacing constraint).
    \item All the wires must not exceed the maximum wirelength $W_{max}$ (max-wirelength constraint).
    \item All timing requirements should be satisfied based on the clocking signal (timing constraint).
    \item Cells of different sizes should not cause perturbations that result in intra-row violations (mixed-cell-size constraint).
\end{itemize}

Therefore, we formulate our objective function for the row-wise AQFP placement problem as follows: 
\begin{equation}
\label{equ:objective}
\begin{split}
    \min_{\textbf{x}} & \quad \sum_{e_i \in E} W(e_i) + \lambda_t T(e_i), \\
    s.t. 
    & \quad \forall e_i \in E \quad W(e_i) \leq W_{max}, \\
    & \quad x_i + w_i \geq x_{i+1} - z_i B, \\
    & \quad x_i + w_i \leq x_{i+1} -z_i s_{min},
\end{split}
\end{equation}
where the $\lambda_t$ is the positive weight for timing cost, $W(e_i)$ is the wirelength cost function of net $e_i$, $W_{max}$ is the maximum allowed wirelength, $z_i$ is a binary value indicating whether two cells are abutted, $B$ is a big positive constant, $s_{min}$ is the minimum spacing, and $w_i$ is the cell width. When $z_i = 0$, it means two cells are abutted, and the two equations correspond to $x_i + w_i = x_{i+1}$. When $z_i = 1$, only $x_i + w_i \leq x_{i+1} - s_{min}$ is required and two cells should maintain a minimal spacing horizontally. $T(e_i)$ is the four-phase timing model cost of net $e_i$, which depends on the located clock phase and is defined as follows:
\begin{equation}
\label{equ:timing}
    T(e_i) = 
    \begin{cases}
    (x_{end} - x_{start}) ^ \alpha,  & \text{if} \ phase \ $\% 4 = 0$\\
    (x_{end} + x_{start}) ^ \alpha,  &  \text{if} \ phase \ $\% 4 = 1$\\
    (- x_{end} + x_{start} ) ^ \alpha,  & \text{if} \ phase \ $\% 4 = 2$\\
    (2\hat{W} - x_{end} - x_{start}) ^ \alpha, & \text{if} \ phase \ $\% 4 = 3$\\
    \end{cases}
\end{equation}
where $x_{start}$ and $x_{end}$ are the start and end $x$ coordinates of net $e_i$, $\hat{W}$ is the layer width of the clock phase, and $\alpha$ is a parameter that modulating the relative importance of the connection across different clock phases, which we set to 2.

\subsubsection{\textbf{Global Placement}}


Global placement aims to determine the best possible cell locations that minimize the overall wirelength while respecting the max-wirelength and spacing constraints of the AQFP. Meanwhile, timing should be addressed during global placement to ensure that the generated placement results do not hold a negative slack. Thus, the objective function should also be timing-aware such that the placement results can work at the desired clock frequency.

A prevalent approach to this constrained minimization problem is to relax the constraints into the objective function and solve the unconstrained minimization problem:
\begin{equation}
\label{equ:objective_unconstrained}
\begin{split}
    \min_{\textbf{x}} & \quad \sum_{e_i \in E} W(e_i) + \lambda_t T(e_i) + \lambda_w (W(e_i) - W_{max} ). \\
\end{split}
\end{equation}
where $\lambda_w$ is the weight for max-wirelength cost.



We utilize DREAMPlace~\cite{lin2019dreamplace} as our analytical global placement engine, incorporating a timing-aware objective function. To address the non-smooth, non-convex nature of the half-perimeter wirelength (HPWL) model, we use a weighted-average (WA) model for more effective wirelength cost optimization.

During global placement, we limit cell location optimization to the one-dimensional $x$ axis, keeping the cell row (clock phase) fixed. Following this, we apply a Tetris-like legalization approach, aiming to preserve global placement results while minimizing cell overspreading. This legalization is iteratively performed to eliminate cell overlaps within each clock phase.

\begin{figure}[htbp]
    \centering
    \includegraphics[width=0.9\linewidth]{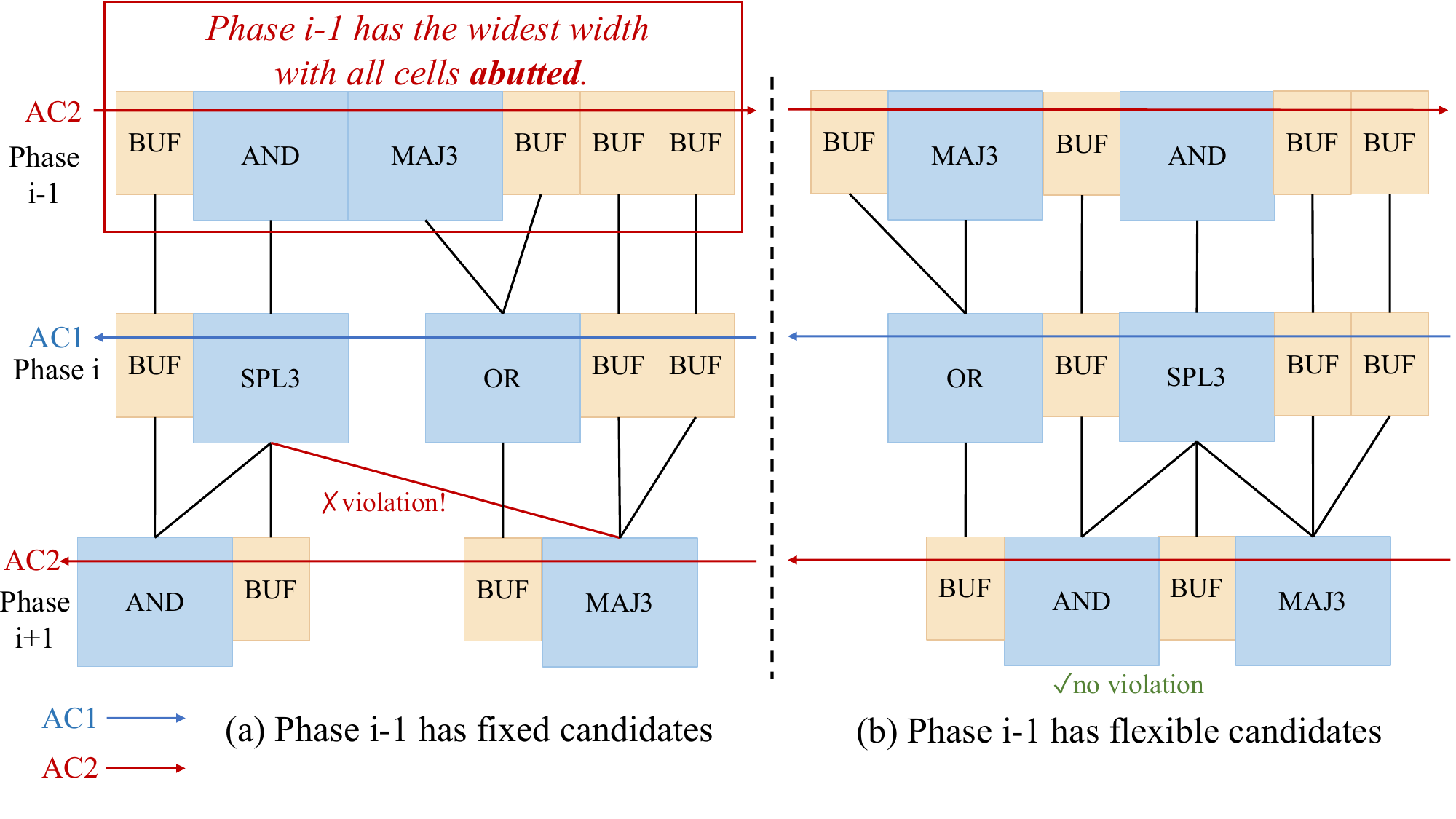}
    \vspace{-0.5cm}
    \caption{\textbf{An illustration of the detailed placement stage where one clock phase has the widest width and all cells are abutted.} (a) When cell candidates are strictly matched to cells of identical sizes, it is possible to end up with a sub-optimal state, leaving some nets with timing violations (highlighted by the red line, SPL3 in phase i to MAJ3 in phase i+1); (b) By allowing flexibility in cell candidates and permitting cell swaps to avoid overlaps (MAJ3, AND, and BUF in phase i-1, in contrast to only MAJ3 and AND in (a)), the detailed placement result is better than (a), exhibiting no timing violations.}
    \label{fig:aqfp_detailed_place}
\end{figure}

\subsubsection{\textbf{Detailed Placement}}


In the detailed placement stage, our goal is to optimally position cells in a non-overlapped circuit netlist $G = (V, E)$ with specified locations $L = (x, y)$, adhering to clocking and mixed-cell-size constraints while minimizing wirelength and timing costs. Since AQFP cells reside in dedicated layers to satisfy the path balancing requirement, a straightforward method is to transform detailed placement to the shortest path problem \cite{dhar2017effective}. This task involves accommodating AQFP's four-phase clocking architecture and varying cell sizes, like 40$\mu$m by 30$\mu$m buffers and 60$\mu$m by 70$\mu$m majority gates. These cells, despite size differences, may coexist in the same clock phase, complicating cell swapping due to proximity issues. As shown in Fig. \ref{fig:aqfp_detailed_place}, if the phase i-1 has the widest width and all cells are abutted, then cell swapping only between candidates of identical sizes may result in sub-optimal solutions. Our framework overcomes this by allowing flexibility in choosing adjacent, non-overlapping cells of different sizes, effectively reducing timing violations in detailed placement.

\subsection{Layer-Wise Fast A* Routing with Space Expansion}

Given a netlist with all placed cell locations, the routing procedure aims to connect all nets with minimized wires and vias following AQFP wiring resources. There are two major differences between AQFP routing and CMOS routing: (\romannumeral1) AQFP uses \texttt{splitter} cells for fan-outs, so the pin connection is one-to-one; (\romannumeral2) the zigzag clocking architecture forces wires to only connect two adjacent layers (clock phases), not across whole chip area. Therefore, a layer-wire routing is sufficient for AQFP, instead of a separate global and detailed router. 


As shown in Algorithm \ref{alg:routing}, we adopt the A* routing algorithm with a priority queue and estimated costs to find the shortest path, incorporating a dynamic step size to limit computation complexity. This means routing wires turn only after a set minimum spacing, for example, 10$\mu$m for the MIT-LL process. While this spacing is efficient, it may cause routability issues in large circuits. To resolve this, we iteratively increase layer distance by the minimum spacing and reroute until all constraints are satisfied. AQFP architecture ensures that this expansion only affects the relative positions of adjacent layers, keeping each layer's placement intact.

\begin{algorithm}[htb!]
\caption{\small Layer-wise A* Routing with Space Expanding}
\label{alg:routing}
\SetAlgoNlRelativeSize{-1}
\SetNlSty{textbf}{\{}{\}}
\SetAlgoNlRelativeSize{-2}
\SetNlSty{}{}{}

\SetKw{KwBreak}{break}
\SetKwFunction{PriorityQueue}{PriorityQueue}
\SetKwFunction{ExpandSpace}{ExpandSpace}
\SetKwFunction{AStar}{A\_star}

\footnotesize
\SetKwInOut{Input}{Input}
\SetKwInOut{Output}{Output}
\SetKwProg{Pn}{Function}{:}{\KwRet{List, cost}}
\Input{Placement results with all cell locations and layers.}
\Output{Routing wires between each layer.}
\Pn{\textsc{\AStar{$graph$, $start$, $goal$}}}{

    let the $Queue$ and $List$ equal an empty list of nodes\;
    put start node on the $Queue$\;

    \While{$Queue$ is not empty}{
        get current node from $Queue$\;
        \If{current node is the $goal$}{
            \KwBreak\;
        }
        update current $cost$\;
        \For{adjacent nodes not in $List$}{
            \If{$cost$ is lower}{
                update current $cost$\;
                put adjacent node on the $Queue$\;
                append current node to $List$\;
            }
        }
    }
}

\ForEach{layer}{
    \While{routing space is not enough}{
        expand spacing between two adjacent layers\;
        \textsc{\AStar{$graph$, $start$, $goal$}}\;
        update whole $graph$ with routed wires\;
    }
}
\end{algorithm}

\subsection{Layout Generation and DRC}

\begin{figure}[htbp]
    \centering
    \includegraphics[width=0.7\linewidth]{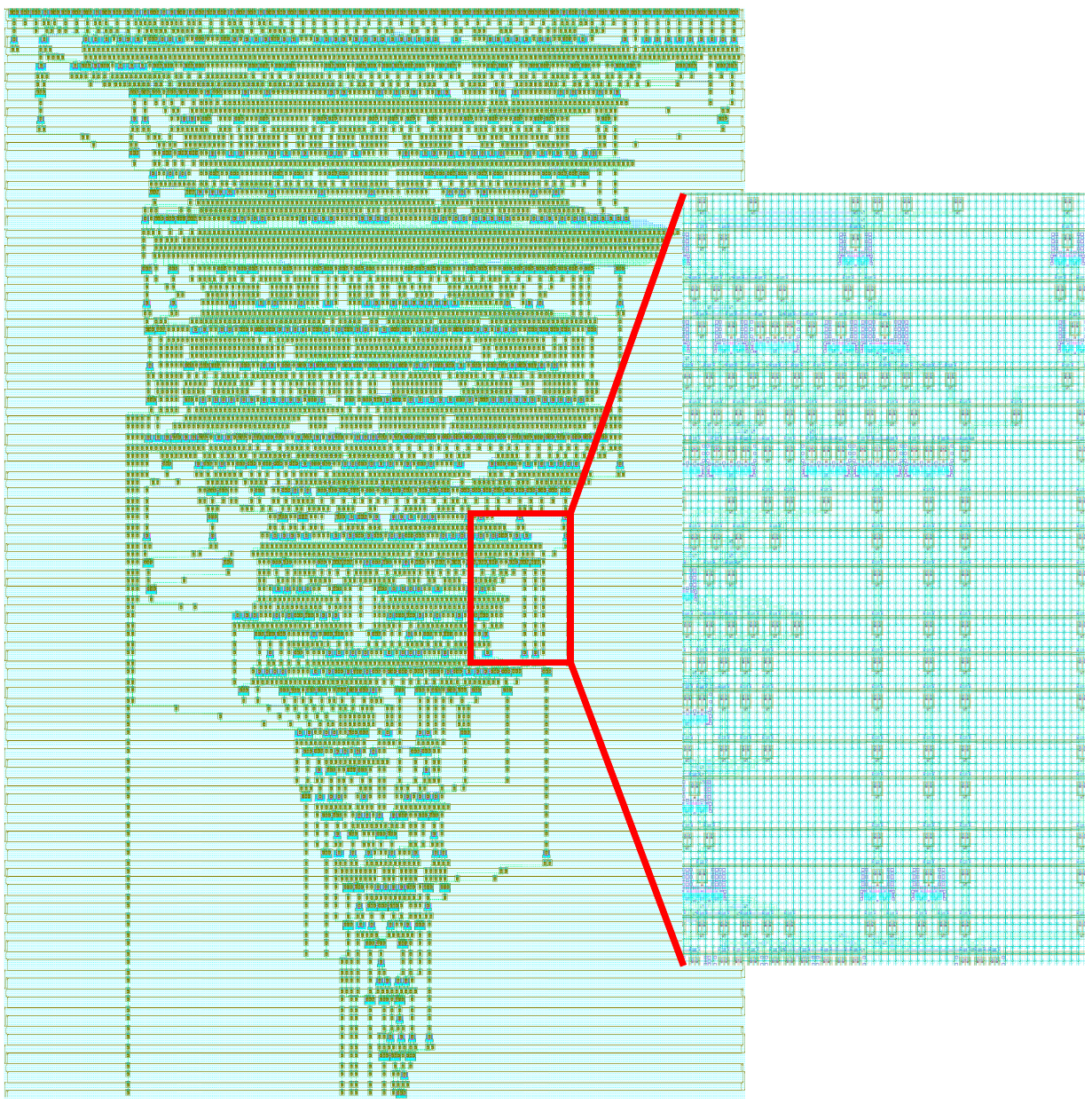}
    \vspace{-0.3cm}
    \caption{\textbf{Layout for AQFP circuits apc128.}}
    \label{fig:aqfp_layout}
\end{figure}

The final step of \N{} is the layout generation and Design Rule Check (DRC).
After placement and routing, we obtain the physical information (coordinates, rotation, wiring paths, track assignment, etc.) for all cells and wires.
Using the AQFP standard cell library, these details are referenced in the layout file, compatible with most layout tools. We leverage KLayout for DRC, checking for rule compliance (e.g., metal layer density, via sizes, contact layer spacing). If the layout passes DRC, we generate the final AQFP GDS layout file; otherwise, we identify and address violation locations, adjusting placement and routing as needed. As an illustration, Fig. \ref{fig:aqfp_layout} displays the completed layout for the AQFP circuit apc128.

\section{Experimental Results}
\label{sec:experiments}

\begin{table}[htbp]
\centering
\ra{1.0} 
\caption{Majority-based logic synthesis results}
\label{tab:statistic}
\resizebox{0.47\columnwidth}{!}{
\begin{tabular}{c|rrr}
\toprule
\textbf{Circuits} & \textbf{\#JJs} & \textbf{\#Nets} & \textbf{\#Delay} \\ \midrule
adder8            &  960           &    462          &  23     \\ \hline
apc32             &  746           &  513            &  21     \\ \hline
apc128            &  5,048         &  2,355          &  45     \\ \hline
decoder           &  2,210         &  989            &  19     \\ \hline
sorter32          &  3,788         &  1,474          &  30     \\ \hline
c432              &  2,500         &  1,048          &  40     \\ \hline
c499              &  4,946         &  2,202          &  31     \\ \hline
c1355             &  4,996         &  2,236          &  31     \\ \hline
c1908             &  4,716         &  2,182          &  34     \\ 
\bottomrule
\end{tabular}
} 
\end{table}


\begin{table*}[htbp]
\begin{center}
\ra{1.0} 
\caption{The comparison on the design performance between GORDIAN-based approach \cite{li2021towards}, TAAS \cite{dong2022taas}, and \N{}.}
\label{tab:placement}
\setlength{\tabcolsep}{1.2pt}
\begin{tabular}{c|rrrrrrrrrrrr}
\toprule

\multirow{2}{*}{\textbf{Circuits}} & \multicolumn{3}{c}{\textbf{GORDIAN-Based \cite{li2021towards}}} & \phantom{abc} & \multicolumn{3}{c}{\textbf{TAAS \cite{dong2022taas}}}  & \phantom{abc}  & \multicolumn{4}{c}{\textbf{\N{}}} \\ \cmidrule{2-4} \cmidrule{6-8} \cmidrule{10-13} 
                & HPWL ($\mu$m) & Buffers & WNS (ps) & & HPWL ($\mu$m) & Buffers & WNS (ps) & & HPWL ($\mu$m) & Buffers & WNS (ps) & Runtime (s) \\ \midrule
adder8          & \bf{10,948}   & 24      & -      & & 12,360      & 24      & -       & & 11,850      & \bf{16}  & -        & 12.1    \\ \hline
apc32           & 15,915        & 26      & -      & & 15,915      & 26      & -       & & \bf{15,530} & 26       & -        & 13.8     \\ \hline
apc128          & 254,068       & 117     & -40.7  & & 245,416     & 110     & -10.1   & & \bf{177,620}& \bf{67}  & \bf{-9.6}& 374.8    \\ \hline
decoder         & \bf{141,151}  & 34      & -8.8   & & 156,213     & \bf{33} & -1.4    & & 153,030     & 43       & \bf{-1.0}& 162.5    \\ \hline
sorter32        & 168,208       & 29      & -6.9   & & 180,427     & 29      & -3.3    & & \bf{132,640}& 29       & \bf{-2.3}& 113.4    \\ \hline
c432            & 51,009        & 46      & -      & & 52,208      & 45      & -       & & \bf{36,050} & \bf{29}  & -        & 50.1     \\ \hline
c499            & 430,658       & 62      & -29.9  & & 431,108     & 62      & -8.9    & & \bf{385,845}& \bf{59}  & \bf{-6.7}& 517.5    \\ \hline
c1355           & 422,556       & 58      & -31.4  & & 426,099     & 58      & -9.1    & & \bf{396,640}& \bf{56}  & \bf{-8.9}& 690.9    \\ \hline
c1908           & 358,271       & 67      & -25.5  & & 361,071     & \bf{66} & -6.9    & & \bf{357,570}&  68      & -6.9     & 353.3    \\ \hline
Average         & 1.112         & 1.178   & 4.045  & & 1.128       & 1.153   & 1.121   & & \bf{1}      & \bf{1}   & \bf{1}   &          \\ 
\bottomrule
\end{tabular}\end{center}
{\footnotesize \raggedright Note: `-' means that the WNS is positive and no timing violations are found under the target clock frequency, which is set as 5GHz. For \N{}, the HPWL for all circuits are all integers of 10$\mu$m, because the AQFP standard cell library goes through an update, and now the cell height, width, pin location, are all integers of 10$\mu$m.  \par}
\end{table*}

\begin{table}[htbp]
\centering
\ra{1.0} 
\caption{Routing Results of \N{}.}
\label{tab:routing}
\resizebox{0.75\columnwidth}{!}{ 
\begin{tabular}{c|rrr}
\toprule
\textbf{Circuits}  & \textbf{\#JJs after Routing} & \textbf{\#Nets} & \textbf{Routed WL ($\mu m$)}\\ \midrule
adder8        &  2,170      & 1,064       & 21,100      \\ \hline
apc32         &  2,040      & 986         & 22,510      \\ \hline
apc128        &  13,860     & 6,761       & 260,770     \\ \hline
decoder       &  7,896      & 3,807       & 252,050     \\ \hline
sorter32      &  8,768      & 3,938       & 218,210     \\ \hline
c432          &  5,286      & 2,531       & 75,710      \\ \hline 
c499          & 19,050      & 9,329       & 816,240     \\ \hline 
c1355         & 21,004      & 10,315      & 932,960     \\ \hline 
c1908         & 15,408      & 7,574       & 617,350     \\ 
\bottomrule
\end{tabular}
} 
\end{table}

\noindent\textbf{Settings:} We implement \N{} in Python and conduct tests on a Linux machine that features an AMD Ryzen Threadripper 2920X processor with 12 cores and 64GB DDR4 memory.

\noindent\textbf{Benchmark Circuits:} We test our framework using classic benchmark circuits for AQFP testing \cite{chen2019adiabatic}, including 8-bit Kogge-Stone adder (adder8), 32-bit/128-bit approximate parallel counter (apc32/apc128), decoder, and 32-bit sorter (sorter32).
We also validate it on the ISCAS’85 benchmark circuits to showcase the effectiveness of our framework \cite{epfl}. Our comprehensive design flow is demonstrated from RTL netlists to final layouts, comparing our results with state-of-the-art AQFP placement tools.

\subsection{Logic Synthesis Results of \N{}}
Table~\ref{tab:statistic} shows the statistic results of AQFP-based netlist after the logic synthesis step. We give the number of JJs, nets, and delay (clock phases) for each AQFP-based netlist. Note that the most basic structure of an AQFP cell (buffer) is a double-Josephson junction, and all other AQFP cells use more than 2 JJs. Therefore, the number of JJs is larger than the number of nets, and the results include all inserted buffers and splitters after the logic synthesis stage.

\subsection{Placement Results of \N{}}

Table~\ref{tab:placement} compares \N{} with the GORDIAN-based approach~\cite{li2021towards} and a timing-aware placer, TAAS~\cite{dong2022taas} for AQFP circuits. We report the HPWL, inserted buffer lines, and the worst negative slack (WNS) using a timing analysis engine~\cite{dong2022taas}.

\N{} shows reduced HPWL and fewer buffers for small circuits like adder8 and apc32. For larger circuits (apc128, c499, c1355, c1908), it achieves significant buffer reductions (up to 42.7\% over GORDIAN-based, 39.1\% over TAAS) and shorter wirelength.
While GORDIAN-based method excels in wirelength, it faces timing issues. TAAS improves timing at a slight wirelength cost. 
This shows that \N{} optimization strategy is effective on large-scale circuits and cell swapping within different-sized cells could save inserted buffer lines.
Overall, \N{} achieves 12.8\% better wirelength with 15.3\% fewer inserted buffer lines and 12.1\% timing improvement over previous state-of-the-art placer, TAAS~\cite{dong2022taas}.

\subsection{Routing Results of \N{}}
Table~\ref{tab:routing} demonstrates the final routing results for AQFP benchmarks including JJs, nets, and routed wirelength. We calculated all the cells and wires in the final layout, including buffers and splitters inserted during the logic synthesis and placement stage.

\section{Conclusions}
\label{sec:conclusion}

In this paper, we present \N{}, a fully-customized RTL-to-GDS design flow specifically tailored for AQFP superconducting circuits. \N{} simultaneously optimizes wirelength and timing while adhering to the clocking and mixed-cell-size constraints inherent in AQFP circuits throughout the design process. In the routing phase, \N{} employs a layer-wise strategy to support space expansion and address potential routability issues. Experimental results demonstrate that \N{} outperforms previous design tools in terms of wirelength and timing for AQFP circuits, setting a robust groundwork for future AQFP applications like RISC-V CPUs and neural network accelerators.~\cite{li2023superbnn}.

\section{Acknowledgment}
This work was supported by JST FOREST Program (Grant Number JPMJFR226W, Japan), and the NSF Expedition program CCF-2124453, NSF CCF-2008514.

\bibliographystyle{IEEEtran}
\bibliography{short_references}

\end{document}